\documentclass[conference]{IEEEtran}
\IEEEoverridecommandlockouts

\usepackage[T1]{fontenc}
\usepackage{cite}
\usepackage{amsmath,amssymb,amsfonts}
\usepackage{graphicx}
\usepackage{booktabs}
\usepackage{xcolor}
\usepackage{url}
\usepackage{balance}
\def\BibTeX{{\rm B\kern-.05em{\sc i\kern-.025em b}\kern-.08em
    T\kern-.1667em\lower.7ex\hbox{E}\kern-.125emX}}

\begin{document}

\title{Partition-Matched Evaluation of Community Features under Distribution
Shift in Android Malware Function-Call Graphs}

\author{
\IEEEauthorblockN{1\textsuperscript{st} Junru Zhu}
\IEEEauthorblockA{\textit{Independent Researcher}\\
Seattle, USA\\
junru.zhuu@gmail.com}
\and
\IEEEauthorblockN{2\textsuperscript{nd} Yixin Yang}
\IEEEauthorblockA{\textit{Independent Researcher}\\
New York, USA\\
vicki\_yang@chiefgroup.com.hk}
\and[\hfill\break\mbox{}\hfill]
\IEEEauthorblockN{3\textsuperscript{rd} Xiaoqing Ding}
\IEEEauthorblockA{\textit{University of Chicago}\\
Chicago, USA\\
alexading@uchicago.edu}
\and
\IEEEauthorblockN{4\textsuperscript{th} Ruoyu Qi}
\IEEEauthorblockA{\textit{Independent Researcher}\\
Charlotte, USA\\
dxfreedom94@gmail.com}}

\maketitle

\begin{abstract}
Graph-based Android malware classifiers can lose accuracy under malware-type
or family shifts. We test whether mesoscopic organization in function-call
graphs provides shift-stable information beyond local degree profiles (LDP),
global statistics, lightweight metadata, and size-matched random partitions.
Using 15,000 MalNet-Tiny, Common, and Distinct graphs, six Leiden descriptors
specified before evaluation, and five optimizer seeds, communities raise Tiny
macro F1 from 78.7\% to 81.3\% but yield 29.4-point source-only Common
degradation. One size-matched random partition yields 27.8-point degradation.
Across five random partitions, mean degradation is 27.9 points; the 95\%
two-level bootstrap interval for random minus community degradation is
$[-4.2,1.2]$ points. With metadata, the corresponding difference is $-0.2$
points with interval $[-1.1,0.6]$. Removing modularity raises structure-only
Common macro F1 from 51.9\% to 53.6\%. The tested signature adds IID signal but
shows no repeatable shift-stability advantage, demonstrating why mesoscopic
graph claims need partition-matched controls and repeated null draws.

\end{abstract}


\section{Introduction}

Android malware classifiers increasingly represent applications as function-call
graphs (FCGs), in which functions are vertices and invocations are directed
edges. This representation supports large structural benchmarks such as
MalNet~\cite{freitas2021malnet}, but its accuracy can deteriorate when malware
types or families change~\cite{tran2026generalization}. Longitudinal Android
studies similarly show that conclusions about robustness depend on behavioral
abstraction and temporally valid evaluation
protocols~\cite{mariconti2017mamadroid,pendlebury2019tesseract}. Understanding
which graph properties remain informative under such shifts is therefore
necessary for separating durable structural evidence from source-specific
correlations.

Community organization is a plausible source of durable structure because
malware functionality is distributed across interacting groups of functions.
The central attribution problem, however, is that common community descriptors
also encode graph scale, connectivity, the number of groups, and their size
distribution. Appending these descriptors can improve a classifier even when
the detected membership has no meaningful alignment with the FCG. A gain over
local or global graph statistics alone therefore cannot establish that actual
community organization supplies the useful information.

We address this attribution problem with a matched random-partition control. For
each application, the control preserves the exact Leiden community-size
multiset while randomizing vertex membership, thereby retaining partition shape
but breaking its alignment with FCG edges. We compute the same six descriptors
for the detected and randomized partitions and compare them alongside a
local-degree-profile (LDP) summary, generic global statistics, and function
metadata. An advantage over both the global and matched-partition controls is
required for a community-specific interpretation.

The evaluation uses 15,000 samples from MalNet-Tiny and its Common
and Distinct variants. Common permits source-only evaluation because it
preserves the source class mapping, whereas Distinct changes the label space and
is evaluated through supervised target adaptation, consistent with the need to
separate graph covariate and concept shifts~\cite{gui2022good}. Adding detected
communities to LDP raises Tiny macro F1 from 78.7\% to 81.3\%, but produces
29.4-point Common degradation compared with 27.8 points for the matched
random-partition control. Across five matched random partitions, mean degradation
is 27.9 points, and the 95\% two-level bootstrap interval for random minus
community degradation is $[-4.2,1.2]$ points. The detected partition is therefore
class-informative without a repeatable community-specific reduction in this
shift.

This study makes three contributions:
\begin{itemize}
    \item a partition-matched protocol that distinguishes edge-aligned community
    information from generic graph and partition-shape effects;
    \item repeated-null evidence, with optimizer-aware uncertainty, that the
    tested descriptor vector adds IID class information but no repeatable
    community-specific robustness on Common; and
    \item descriptor-shift, ablation, and class-level analyses that identify
    modularity-associated heterogeneity behind the aggregate result.
\end{itemize}

\section{Related Work}

\subsection{Malware representations under distribution shift}

Android malware detection has progressed from sparse, explainable application
features~\cite{arp2014drebin} and deep classifiers~\cite{zhang2018android} to
behavioral abstractions and graph representations. MaMaDroid summarizes
API-call sequences as package- or family-level transition models to reduce
sensitivity to evolving APIs~\cite{mariconti2017mamadroid}, while MalNet enables
large-scale classification with application FCGs~\cite{freitas2021malnet}.
The FCG literature also includes whole-graph embeddings for detection and
family classification~\cite{xu2021detecting}, deep message passing with jumping
knowledge~\cite{lo2022jumping}, and heterogeneous app--API graph
learning~\cite{gao2021gdroid}. These studies establish the predictive value of
graph representations but do not test whether classifier gains attributed to a
detected partition survive a partition-shape-matched null. The
closest OOD benchmark to our setting augments MalNet graphs with function
metadata and code embeddings and evaluates the Common and Distinct
shifts~\cite{tran2026generalization}. These representations test which semantic
and structural signals transfer, but the benchmark does not isolate whether a
community descriptor reflects edge-aligned organization or merely the shape of
a graph partition.

\subsection{Community-based graph representation}

Community analysis has also been used for Android malware attribution.
Andro-Simnet detects groups in an application-similarity network, where each
vertex represents an entire application~\cite{kim2018androsimnet}; our setting
instead studies groups of functions within each application's FCG. At this
scale, community methods span disjoint and overlapping
formulations~\cite{fortunato2010community}. Clique-based approaches expose
overlap directly~\cite{palla2005overlap}, and recent hybrid methods combine
clique structure with structural-feature node extraction to preserve
connectivity~\cite{ma2026hybrid}. We study a complementary attribution question
for a disjoint partition. Modularity quantifies excess within-group
connectivity relative to a degree-based null model~\cite{newman2006modularity},
and Leiden produces well-connected partitions through iterative
refinement~\cite{traag2019leiden}. Descriptors computed from such partitions
nevertheless retain the number and sizes of the detected groups. Our matched
random-partition control preserves these quantities while breaking their
alignment with FCG edges, enabling a community-specific rather than
partition-shape interpretation.

\section{Method}

\subsection{Overview and graph preprocessing}

Given an application graph $G=(V,E)$, our pipeline constructs local and global
structural controls, detects a Leiden partition, generates a random partition
with the same community-size multiset, and maps both partitions to the same
descriptor vector. Function metadata can then be concatenated before
classification. The matched control isolates the central attribution question:
an advantage for the detected partition over its size-matched counterpart
requires information from alignment between communities and FCG edges, rather
than partition shape alone.

We remove self-loops, merge parallel edges, symmetrize the directed invocation
graph for community detection, and remove isolated vertices. Labels and the
released train, validation, and test indices are unchanged. Preprocessing is
deterministic and cached by dataset digest, graph index, algorithm, resolution,
and seed.

\subsection{Local and global structural controls}

The LDP of a vertex consists of its degree and the minimum, maximum, mean, and
standard deviation of its neighbors' degrees. To obtain a fixed-size graph
vector, we summarize each of these five channels by its mean, standard
deviation, first quartile, median, third quartile, and maximum, yielding 30
features~\cite{cai2018ldp}. LDP therefore controls for local degree
organization without computing a graph partition.

The generic global control contains six descriptors chosen before evaluation:
$\log(1+|V|)$, $\log(1+|E|)$, density, degree coefficient of variation,
normalized connected-component size entropy, and the largest-component ratio.
These features capture graph scale, sparsity, heterogeneity, and fragmentation
without using a community partition.

\subsection{Detected-community descriptor vector}

We run Leiden with the modularity objective and resolution 1.0 on each
preprocessed FCG. Let the resulting partition be
$\mathcal{C}=\{C_1,\ldots,C_k\}$. To summarize both partition shape and its
alignment with the graph, we compute
\begin{equation}
\begin{split}
\phi_{\mathrm{com}}(G)=\bigg[
&\log(1+k), Q, H_C, r_{\max},\\
&\log\left(1+\frac{m_{\mathrm{in}}}{m_{\mathrm{out}}+1}\right), r_B
\bigg].
\end{split}
\end{equation}
Here, $Q$ denotes modularity, $H_C$ is community-size entropy normalized by
$\log k$, and $r_{\max}=\max_i |C_i|/|V|$. The variables $m_{\mathrm{in}}$ and
$m_{\mathrm{out}}$ count edges within and across communities, while $r_B$ is
the fraction of vertices incident to at least one inter-community edge.
Community count, size entropy, and largest-community ratio describe partition
shape. Modularity, the intra/inter-community edge ratio, and the bridge-node
ratio measure how that partition aligns with FCG edges.

\subsection{Matched random-partition control}

For every graph, we keep the multiset
$\{|C_1|,\ldots,|C_k|\}$ fixed, permute vertices with a deterministic recorded
seed, and assign consecutive blocks to synthetic communities. We then compute
the same descriptor vector. Its first, third, and fourth components are
identical to those of the detected partition by construction; only modularity,
the intra/inter-community edge ratio, and the bridge-node ratio can change.
Thus, comparisons with this control hold partition shape fixed and test whether
edge-aligned community organization contributes predictive information.
The main comparison uses one seed specified before model evaluation. A
sensitivity analysis repeats the construction with four additional independent
seeds, giving five matched partitions per graph.

\subsection{Function metadata and classifier}

The benchmark provides a 201-channel metadata tensor for each function. We
aggregate every channel by its graph-wise mean, standard deviation, and maximum,
yielding 603 graph-level metadata features. These experiments use only the
released tensor and do not access APKs, decompiled source, or executable
artifacts.

Each representation is standardized using Tiny training statistics and supplied
to a two-layer multilayer perceptron with 64 hidden units per layer and a
five-class output head. AdamW uses learning rate $10^{-3}$, weight decay
$10^{-4}$, and batches of 128. Training runs for at most 160 epochs with
checkpoint selection by validation macro F1, patience 24, and no test-time model
selection. We run five pre-specified optimizer seeds.

\subsection{Supervised target adaptation}

The Tiny-trained model is evaluated unchanged on Tiny test and Common test. For
Distinct, the encoder is initialized from the Tiny model, the output head is
reinitialized, and all parameters are adapted for at most 80 epochs on the
Distinct training split. A same-budget randomly initialized model is the
supervised scratch control. Distinct features retain the Tiny scaler so that
adaptation does not use target-distribution normalization.

\section{Experimental Setup}

\subsection{Data, shifts, and metrics}

We use the released MalNet-Tiny shift benchmark
\cite{freitas2021malnet,tran2026generalization}. Tiny, Common, and Distinct each
contain 5,000 graphs, divided into 3,500 training, 500 validation, and 1,000 test
examples. Tiny and Common share a family-level class mapping but contain
different malware types, so the Tiny model is evaluated on Common without
retraining. Distinct changes both families and types and is evaluated after
labeled target training. Common therefore measures source-only transfer under a
shared label space, whereas Distinct measures supervised target adaptation.

We report accuracy, macro F1, weighted F1, and one-vs-rest macro AUROC, with
macro F1 as the primary metric. Values are means and sample standard deviations
over five optimizer seeds. For Common, lower OOD degradation is better:
\begin{equation}
\Delta_{\mathrm{Common}} =
F_{1,\mathrm{macro}}^{\mathrm{Tiny}} -
F_{1,\mathrm{macro}}^{\mathrm{Common}}.
\end{equation}

\subsection{Comparison protocol}

The primary attribution test compares LDP plus the detected-community descriptor
vector with LDP alone, LDP plus global statistics, and LDP plus the matched
random-partition control. Evidence for edge-aligned community information
requires lower mean Common degradation than both the global and matched
random-partition controls, using paired optimizer seeds. The secondary test
repeats this decision rule after concatenating LDP and function metadata.
To test sensitivity to null-partition sampling, we average over five matched
partition draws and five optimizer seeds. We report a 10,000-replicate
two-level bootstrap interval for random minus detected-community degradation,
resampling partition draws and optimizer seeds independently; positive values
favor detected communities.

Mechanism analyses remove modularity or the bridge-node ratio, inspect
class-level F1, and measure each descriptor's Tiny--Common shift using
1-Wasserstein distance normalized by its pooled standard deviation. For
Distinct, source-initialized adaptation is compared with target-scratch training
under the same budget. The 60-model repeated-null grid retains 120 metric
records and 120,000 test predictions. Feature audits reproduce the primary
community and random features and verify partition-shape preservation for every
null draw.

\section{Results}
\subsection{Detected communities add IID signal but not partition-specific stability}

Table~\ref{tab:main} and Fig.~\ref{fig:results}(a) separate source predictiveness
from shift stability. LDP obtains $78.7\pm1.0$ macro F1 on Tiny and
$52.2\pm3.3$ on Common, corresponding to $26.4\pm3.8$ points of degradation.
Adding the detected-community descriptor vector raises Tiny macro F1 by 2.6
points to $81.3\pm0.9$, while Common macro F1 is $51.9\pm1.9$. The resulting
29.4-point degradation shows that the added IID signal does not transfer to
Common.

The controls identify which part of this gain is community-specific. LDP plus
global statistics has 30.2-point degradation, so detected communities improve
over the generic global control by 0.9 points on average. The pre-specified
matched random-partition control instead reaches $52.7\pm2.5$ Common macro F1 and
27.8-point degradation, 1.6 points below the detected partition. Detected
communities have lower degradation in three of five paired seeds, but the two
opposite differences are larger. The primary attribution criterion is therefore
not met: the detected partition is source-predictive without a reproducible
advantage over the same partition shape with randomized membership.

\begin{table*}[t]
\caption{Performance in percent (mean $\pm$ standard deviation over five
seeds). Degradation is Tiny minus Common macro F1; lower is better. Detected
communities do not outperform matched random-partition controls on degradation.}
\label{tab:main}
\centering
\footnotesize
\resizebox{\textwidth}{!}{\begin{tabular}{lccccc}
\toprule
Representation & Tiny macro F1 & Common acc. & Common macro F1 & Common weighted F1 & Degradation \\
\midrule
LDP summary & 78.7 $\pm$ 1.0 & 51.4 $\pm$ 3.4 & 52.2 $\pm$ 3.3 & 52.2 $\pm$ 3.3 & 26.4 $\pm$ 3.8 \\
LDP + global & 80.5 $\pm$ 0.9 & 49.5 $\pm$ 2.2 & 50.2 $\pm$ 2.3 & 50.2 $\pm$ 2.3 & 30.2 $\pm$ 2.4 \\
LDP + matched random & 80.5 $\pm$ 0.5 & 52.0 $\pm$ 2.6 & 52.7 $\pm$ 2.5 & 52.7 $\pm$ 2.5 & 27.8 $\pm$ 2.7 \\
LDP + community & 81.3 $\pm$ 0.9 & 51.4 $\pm$ 1.8 & 51.9 $\pm$ 1.9 & 51.9 $\pm$ 1.9 & 29.4 $\pm$ 1.7 \\
LDP + community, no modularity & 81.6 $\pm$ 0.9 & 53.0 $\pm$ 2.3 & 53.6 $\pm$ 2.4 & 53.6 $\pm$ 2.4 & 28.0 $\pm$ 2.4 \\
Metadata & 91.5 $\pm$ 0.5 & 53.7 $\pm$ 0.8 & 51.6 $\pm$ 1.4 & 51.6 $\pm$ 1.4 & 39.9 $\pm$ 1.4 \\
Metadata + global & 91.5 $\pm$ 0.4 & 54.5 $\pm$ 1.1 & 52.2 $\pm$ 1.6 & 52.2 $\pm$ 1.6 & 39.2 $\pm$ 1.5 \\
Metadata + community & 92.0 $\pm$ 0.2 & 54.0 $\pm$ 1.0 & 51.5 $\pm$ 1.2 & 51.5 $\pm$ 1.2 & 40.4 $\pm$ 1.3 \\
LDP + metadata & 91.4 $\pm$ 0.3 & 53.6 $\pm$ 1.3 & 51.0 $\pm$ 1.1 & 51.0 $\pm$ 1.1 & 40.4 $\pm$ 0.9 \\
LDP + metadata + global & 91.4 $\pm$ 0.7 & 53.6 $\pm$ 1.1 & 51.1 $\pm$ 1.1 & 51.1 $\pm$ 1.1 & 40.3 $\pm$ 0.8 \\
LDP + metadata + matched random & 91.4 $\pm$ 0.3 & 54.5 $\pm$ 0.7 & 52.4 $\pm$ 1.1 & 52.4 $\pm$ 1.1 & 39.0 $\pm$ 1.2 \\
LDP + metadata + community & 91.2 $\pm$ 0.4 & 53.8 $\pm$ 0.7 & 51.3 $\pm$ 0.8 & 51.3 $\pm$ 0.8 & 40.0 $\pm$ 0.6 \\
\bottomrule
\end{tabular}
}
\end{table*}

\subsection{Repeated null partitions do not reveal a community advantage}

Table~\ref{tab:null-robustness} tests whether the main comparison depends on one
random partition. For LDP, detected-community degradation is 29.4 points,
whereas the mean over five matched partitions and five optimizer seeds is 27.9
points. Random minus community degradation is $-1.5$ points with a 95\%
two-level bootstrap interval of $[-4.2,1.2]$. Each partition-draw mean lies
between 27.5 and 28.1 points, below the detected partition's 29.4 points.

With metadata, detected and random degradation are 40.0 and 39.8 points. Their
$-0.2$-point difference has interval $[-1.1,0.6]$. Optimizer variation therefore
precludes a strict superiority claim, but repeated null sampling does not
produce a stable community-specific advantage in either representation.

Structure-only degradation varies by 0.7 points across partition draws but by
4.6 points across optimizer-conditioned random means. After averaging draws,
the paired random-minus-community differences are $+1.7$, $-5.1$, $+1.8$,
$-1.0$, and $-4.9$ points: detected communities win two seeds and random
membership wins three. Optimizer resampling is therefore retained in the
reported interval.

\begin{table}[!t]
\caption{Repeated matched-partition sensitivity. Degradation is in macro-F1
points; lower is better. Random values average five partition draws and five
optimizer seeds. Brackets give a 95\% two-level bootstrap interval for random
minus community; positive favors detected communities.}
\label{tab:null-robustness}
\centering
\footnotesize
\resizebox{\columnwidth}{!}{\begin{tabular}{lccc}
\toprule
Representation & Community deg. & Random deg. & Random $-$ community \\
\midrule
LDP & 29.4 & 27.9 & -1.5 [-4.2, 1.2] \\
LDP + metadata & 40.0 & 39.8 & -0.2 [-1.1, 0.6] \\
\bottomrule
\end{tabular}
}
\end{table}

\subsection{Matched randomization removes edge alignment}

Table~\ref{tab:control-validity} verifies the control operation. Shape
descriptors remain identical, while median modularity falls from 0.690 to
$-0.004$ on Tiny and from 0.711 to $-0.002$ on Common. The log intra/inter edge
ratio falls from about 1.6 to below 0.08, and the bridge ratio rises from about
0.24 to above 0.97. Randomization therefore disrupts edge alignment on both
domains.

\begin{table}[!t]
\caption{Median edge-alignment descriptors. For matched random partitions, the
five draw values are averaged per graph before taking the dataset median. Shape
descriptors are identical by construction and omitted.}
\label{tab:control-validity}
\centering
\footnotesize
\resizebox{\columnwidth}{!}{\begin{tabular}{llrrr}
\toprule
Dataset & Partition & $Q$ & Log ratio & Bridge ratio \\
\midrule
Tiny & Detected & 0.690 & 1.609 & 0.235 \\
Tiny & Matched random & -0.004 & 0.073 & 0.970 \\
Common & Detected & 0.711 & 1.625 & 0.251 \\
Common & Matched random & -0.002 & 0.059 & 0.977 \\
\bottomrule
\end{tabular}
}
\end{table}

\subsection{Modularity concentrates class-specific instability}

Removing the bridge-node ratio changes Common macro F1 only from
$51.9\pm1.9$ to $51.8\pm1.8$. Removing modularity instead raises Common macro
F1 to $53.6\pm2.4$, the strongest structure-only Common result, while retaining
$81.6\pm0.9$ on Tiny. Its 28.0-point degradation recovers most of the gap to the
matched random-partition control, although it remains 0.2 points higher.

The class-level effects localize this instability. Relative to LDP, the full
community vector raises addisplay/dowgin F1 from 43.6\% to 48.5\% but lowers
adware/startapp from 47.6\% to 38.9\%. Removing modularity restores
adware/startapp to 46.8\% and raises benign F1 from 48.8\% to 52.5\%.
The effect is label-dependent. Modularity accounts for most of the adverse
shift observed for adware/startapp.

\subsection{Marginal descriptor shift does not determine predictive stability}

\begin{figure*}[t]
\centering
\begin{minipage}[t]{0.48\textwidth}
\vspace{0pt}
\centering
\IfFileExists{figures/main_results.pdf}{
\includegraphics[width=\linewidth]{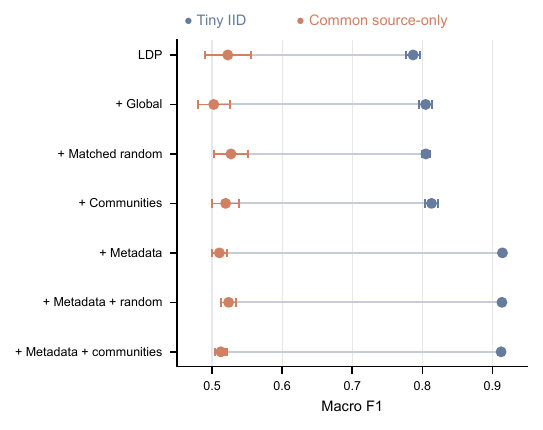}
}{}
\textbf{(a)}
\end{minipage}
\hfill
\begin{minipage}[t]{0.48\textwidth}
\vspace{0pt}
\centering
\IfFileExists{figures/feature_stability.pdf}{
\includegraphics[width=\linewidth]{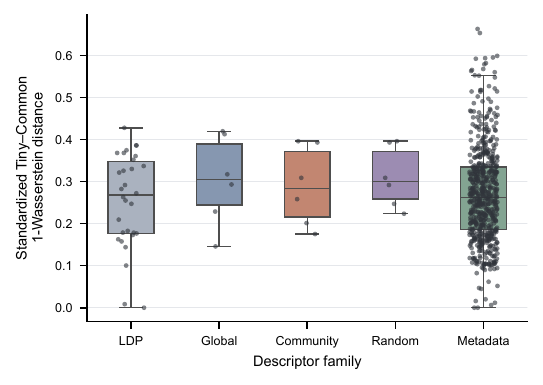}
}{}
\textbf{(b)}
\end{minipage}
\caption{Common-shift evidence. (a) Tiny IID and Common source-only macro F1
(mean $\pm$ standard deviation over five seeds); the pre-specified matched
random partition preserves each graph's Leiden community-size multiset.
(b) Standardized
Tiny--Common 1-Wasserstein shift by descriptor family. Marginal stability does
not imply partition-specific classification gain.}
\label{fig:results}
\end{figure*}

Fig.~\ref{fig:results}(b) measures each descriptor's Tiny--Common movement.
Median standardized 1-Wasserstein distance is 0.268 for LDP, 0.305 for global
statistics, 0.284 for community descriptors, 0.300 for matched random
descriptors, and 0.262 for metadata. Community descriptors move less than the
global and random descriptors in isolation, yet they do not provide the lowest
classification degradation. Marginal descriptor stability is therefore useful
as a diagnostic but does not determine partition-specific predictive stability.

\subsection{Metadata preserves IID accuracy but not a community advantage}

Metadata alone obtains $91.5\pm0.5$ macro F1 on Tiny and $51.6\pm1.4$ on
Common, a 39.9-point degradation. Adding detected communities raises Tiny to
$92.0\pm0.2$ but gives $51.5\pm1.2$ on Common and 40.4-point degradation.

The LDP-plus-metadata comparison repeats the control ordering. Its degradation
is 40.4 points without additional graph descriptors, 40.3 with global
statistics, 40.0 with detected communities, and 39.0 with the matched random
partition. Detected communities improve over the corresponding global control
by 0.4 points but remain 1.0 point above the matched control. Function metadata
therefore preserves high IID accuracy without creating a community-specific
Common advantage.

\subsection{Source initialization has little effect under Distinct supervision}

Under the same labeled target budget, most source-initialized versus
target-scratch differences are below 0.4 macro-F1 points. LDP plus communities
obtains $94.7\pm0.2$ from source initialization and $94.7\pm0.4$ from scratch.
The largest difference is 0.85 points for LDP plus global and community
features.

\section{Discussion}
Detected communities improve Tiny but show no repeatable Common-degradation
advantage over matched random controls, so class-correlated source signal does
not imply edge-aligned shift stability. All five structure-only null-draw means
favor random membership, although the optimizer-aware interval crosses zero.
The result is therefore evidence against a stable advantage, not a claim of
equivalence or significant random-partition superiority. The control preserves
the community-size multiset while randomizing membership, isolating
membership--edge alignment.

Modularity accounts for most of the observed instability: removing it recovers
most of the gap and changes class-level behavior, whereas removing the
bridge-node ratio has little effect. Metadata does not alter the control
ordering, and low marginal descriptor shift coexists with large Common
degradation. Marginal feature movement is therefore insufficient to infer
classifier stability.

Partition-derived features should be compared with null partitions that
preserve classifier-visible partition statistics. Here, the matched control
and repeated draws change the interpretation obtained from local and global
controls alone: the IID gain is not specific to detected organization. This
conclusion applies to six whole-graph descriptors; community-conditioned
function representations remain a separate design space.

\textbf{Implications for representation design.} The descriptor vector contains
three partition-shape quantities that are identical under the matched control
and three quantities that change with membership--edge alignment. Because
random membership remains competitive after repeated draws, the combined
vector acts more like a generic transform of graph topology than a stable
summary of functional modules. Stronger designs could pool function metadata
within communities, represent typed cross-community calls, or aggregate
partitions across resolutions. They would still need repeated matched controls:
target accuracy alone cannot separate a transferable community mechanism from
another source-correlated statistic. Tiny--Common degradation must therefore be
reported alongside target macro F1.

Operationally, a community-aware representation should be reported as a
three-part test: source IID lift, source-only degradation under a shared label
space, and the paired gap to repeated matched nulls. The last quantity should
resample both partitions and optimizer seeds. This order matters: selecting by
Tiny gain alone favors source-correlated descriptors, while selecting by Common
target F1 alone can hide a lower source baseline. Reporting all three quantities
makes the attribution claim falsifiable and distinguishes improved
classification from evidence that detected membership is responsible.

\section{Threats to Validity}

\textbf{Benchmark scope.} We evaluate one five-label MalNet-derived benchmark.
Common preserves malware families while shifting malware type, so it does not
characterize temporal or cross-corpus evolution. Distinct changes the label
space and uses supervised target data, measuring adaptation rather than
source-only transfer.

\textbf{Graph and partition construction.} Static FCGs may omit reflective and
dynamically loaded calls. Symmetrization removes call direction before Leiden
detection, restricting the community result to undirected partitions. We use
one modularity objective, resolution, and Leiden seed, so the study does not
measure sensitivity to alternative objectives, resolutions, or partition
initializations. The matched null preserves community sizes but not the
within-community degree distribution or connectedness. Degree-stratified
assignment and constrained edge-swap nulls could separate membership semantics
from degree allocation more tightly.

\textbf{Representation scope.} We evaluate six graph-level community
descriptors with an MLP; conclusions do not extend to node-level or
community-conditioned message passing. The matched control isolates partition
shape from edge alignment but does not establish correspondence between
detected communities and functional software modules.

\textbf{Statistical scope.} Five optimizer seeds and five matched-partition
draws estimate optimization and null-sampling variation on one released split.
The two-level bootstrap does not cover variation across data splits or
community-detection settings, and an interval crossing zero is not evidence of
equivalence.

\section{Conclusion}
Whole-graph community descriptors encode IID signal in Android malware FCGs
but show no repeatable Common-shift stability advantage over size-matched random
partitions, with or without metadata. The modularity ablation localizes much of
the class-specific gap to one partition-alignment descriptor. Repeated
matched-partition controls separate partition shape from edge-aligned
organization and expose uncertainty that a single null draw cannot, making
community-specific robustness claims directly testable.

\balance
\bibliographystyle{IEEEtran}
\bibliography{references}

\end{document}